\documentclass[prl,aps,twocolumn,reprint,floatfix,showpacs,10pt]{revtex4-1}
\usepackage{amsmath}
\usepackage{graphicx}
\usepackage{epsfig}
\usepackage{graphics}
\usepackage{bm}
\usepackage{amssymb}
\usepackage{psfrag}
\usepackage{mathtools}
\usepackage[english]{babel}
\usepackage[T1]{fontenc}
\usepackage{color}
\usepackage[dvipsnames]{xcolor}
\usepackage{braket}
\usepackage{ulem}
\usepackage{umoline}

\usepackage[colorlinks,bookmarks=false,citecolor=blue,linkcolor=red,urlcolor=black]{hyperref}
\usepackage{amssymb,amsmath} 
\usepackage[T1]{fontenc}
\usepackage{epsfig}

\def\urlprefix{}
\def\url#1{}
\def\be{\begin{equation}}
\def\ee{\end{equation}}
\def\bea{\begin{eqnarray}}
\def\eea{\end{eqnarray}}
\def\bi{\begin{itemize}}
\def\ei{\end{itemize}}
\def\bin{\begin{enumerate}}
\def\ein{\end{enumerate}}

\def\bg{\begin{equation}\begin{gathered}}
\def\eg{\end{gathered}\end{equation}}

\def\bgg{\begin{equation*}\begin{gathered}}
\def\egg{\end{gathered}\end{equation*}}

\def\bgg{\begin{equation*}\begin{gathered}}
\def\egg{\end{gathered}\end{equation*}}
\def\B#1{\!\left(#1\right)}

\begin{document}
\title{Many-body localization due to random interactions}
\author{Piotr Sierant$^{1}$, Dominique Delande$^{2}$,  and Jakub Zakrzewski$^{1,3}$\email[email to:]{jakub.zakrzewski@uj.edu.pl}}
\affiliation{ \mbox{$^1$ Instytut Fizyki imienia Mariana
Smoluchowskiego, Uniwersytet Jagiello\'nski, ulica \L{}ojasiewicza
11, PL-30-348 Krak\'ow, Poland.} 
\mbox{$^2$Laboratoire Kastler Brossel, UPMC-Sorbonne
Universit\'es, CNRS, ENS-PSL Research University,}
\mbox{Coll\`ege de France; 4
Place Jussieu, 75005 Paris, France} 
\mbox{$^3$ Mark Kac Complex
Systems Research Center, Uniwersytet Jagiello\'nski, Krak\'ow,
Poland. }}

\date{\today}

\begin{abstract}
The possibility of  observing many body localization of ultracold atoms in a one dimensional optical lattice
is discussed for random interactions. In the non-interacting limit, such a system reduces to single-particle physics 
in the absence of disorder, i.e. to  extended states. 
In effect the observed localization is inherently due to interactions and is thus a genuine many-body effect.
In the system studied, many-body localization manifests itself in a lack of thermalization visible in temporal 
propagation of a specially prepared initial state,  in transport properties, in the logarithmic growth of entanglement entropy
as well as in statistical properties of energy levels.
\end{abstract}
%\date{\filemodprint{\jobname}~~File: \textbf{\jobname}}
\maketitle

Almost 60 years ago Anderson \cite{Anderson58} has shown that  disorder has dramatic 
effects on properties of noninteracting particles. For one-dimensional (1D) systems, even the smallest 
disorder generically leads to exponential localization of eigenfunctions killing any long-range transport. 
This is in a
striking contrast with orderly periodic structures supporting Bloch waves as eigenfunctions. 
Interestingly, the original idea of Anderson was to consider the effect of disorder on interacting particles. This more difficult problem is not fully understood up till now. Anderson localization is often described as the result of quantum mechanical interference of different multiple scattering paths. This interpretation sheds some light on the interacting particles problem. In the sequel of this Letter, we discuss the physics of ultra-cold atoms in a 1D optical lattice, in conditions where decoherence is negligible on the time scale
of the (numerical) experiment.
For weak 
interactions and small disorder, an effective mean field description of the 
system is possible leading to the Gross-Pitaevski equation.
The latter, however, is a nonlinear equation, for which no superposition principle works;
the concept of interference cannot be easily applied \cite{Flach11}. Even for just two interacting particles 
it was shown that the localization length rapidly grows with the strength of the interactions \cite{Dima94}. 
This makes extremely difficult to observe Anderson localization in the presence of interactions \cite{Schulte05,Clement05,Fort05}.
This was the primary reason why the cold-atom observations of Anderson localization were carried out in the noninteracting regime \cite{Billy08,Roati08}.

A novel path in the investigation of localization was initiated 
in the paper of Basko, Aleiner and Altschuler \cite{Basko06} where,
using a  perturbative approach, it was shown that  there may exist a 
transition to localized states for a sufficiently strong disorder. 
In such a situation, the mean field approach is not applicable and the
full many body quantum theory has to be used. The latter is linear and 
the superposition principle holds:  the picture of interfering paths is restored. 
There is, however, an another conceptual problem: in which sense one
may speak of localization? A possible answer is that one should no longer 
consider the configuration space but rather think in terms of localization in Hilbert space \cite{Huse14}.

Many-body localization (MBL) became recently a hot topic
(a search in arXiv for ``many-body localization'' in the title or abstract yields more than 150 papers in the last 12 months).  
MBL is very  often connected with a lack
of thermalization in the system. While the whole isolated system evolves in a 
fully coherent way, one may ask whether a small subsystem  shows signs of 
thermalization, \textit{i.e.} whether the system evolves in such a way that memory about the initial state is eventually lost
 \cite{Gogolin15}. 
Often lack of thermalization (in the sense of averages of observables) is assumed
as a very definition of MBL \cite{Huse14}. 

Some phenomenological understanding of MBL  can be obtained using an effective 
integrability approach \cite{Serbyn13a,Serbyn13b} or using the renormalization 
group approaches \cite{Vosk13,Vosk14}. Most of the treatments, however, 
are numerical, primarily related to spin-1/2 chains, e.g. Heisenberg 
model with a random magnetic field \cite{Znidaric08} or the XX model
(e.g. \cite{Bardarson12}). That is due to the complexity of many body problems:
spin 1/2 chains may be efficiently treated numerically using time-dependent Density
Matrix Renormalization Group (tDMRG) methods
~\cite{Schollwoeck11}. Much less often can one meet simulations for cold-atomic 
systems, see~\cite{Tang15} and references therein.

An attempt to observe MBL in a 1D system has been reported in~\cite{Schreiber15}, where a
 system of interacting fermions in an optical lattice potential (effectively one-dimensional) is studied. 
 The initial state is carefully prepared in such a way that a single fermion  occupies every second site, 
 other sites being empty. During the temporal evolution in the absence of disorder, the occupations of 
 different sites equalize on average - the system ``thermalizes''. The addition of 
a sufficiently strong quasi-random disorder (adding a second lattice with a period incommensurate with 
the primary lattice) allows one to observe a novel behavior: a partial asymmetry of occupations of odd 
and even sites survives for intermediate times pointing towards a lack of thermalization. This is taken 
as a signature of MBL. The study is supplemented with simulations using tDMRG that reveal a logarithmic 
in time increase of the entropy of entanglement during the course of evolution: this is another possible 
signature of MBL \cite{Znidaric08,Bardarson12,Serbyn13a,Rahul15}. 
However, the localization/delocalization transition takes place at a disorder strength very close to 
the threshold for single particle localization in the Aubry-Andr\'e {model}~\cite{Aubry80}, and additionally only weakly depends
on the interaction strength. The MBL observed thus has a predominantly single-particle character and
it is a perturbation of the single particle physics \cite{Basko06}. 

The aim of this letter is to show that ultracold atoms allow us to study models with genuine nonperturbative MBL.
One possible way  is to consider a system that does not show localization in the absence of interactions.
This rules out single particle localization mechanisms. Consider bosons or fermions in a regular optical lattice in
the absence of interactions. This is 
a perfectly regular system described by Bloch waves as eigenstates. Now we turn on -- in a controlled way -- 
interactions that randomly depend on position. Such a situation may be realized close to a Feshbach resonance when the scattering length 
strongly depends on the magnetic field value. 
If the latter fluctuates in space, a system with the desired properties is created.
Various phases of this model were studied {in} \cite{Gimperlein05}. In particular, it was shown that 
the bosonic Mott insulator entirely disappears for sufficiently large occupations in the system. While the ground state 
was considered in \cite{Gimperlein05}, we analyze here the properties of excited states of the system inspecting their 
localization properties. We shall study eigenvalue statistics for systems of small size ({allowing} a partial comparison with 
\cite{Tang15,Serbyn16,Garcia-Garcia}) as
well as the time propagation (using tDMRG) of appropriately prepared initial states (as in \cite{Schreiber15}).

The Bose-Hubbard Hamiltonian describing a 1D system in an optical lattice within the 
tight binding approximation reads, assuming random on-site interactions 
\bg
\label{H}
\hat{H}= -J \sum_{i}^{L-1}\B{\hat{a}_{i+1}^\dag \hat{a}_i + {\rm h. c.}}
+\frac{1}{2} \sum_{i}U_i\hat{n}_i\B{\hat{n}_i-1}, \\
[\hat{a}_i,\hat{a}_j^\dag]=\delta_{ij}, \ [\hat{a}_i,\hat{a}_j]=0, 
\  \hat n_i=\hat{a}_i^\dag\hat{a}_i,
\eg
with the first term describing the tunneling while the second term corresponds to interactions. 
Here, following \cite{Gimperlein05} we assume the interaction strength to depend on site taking $U_i=Ux_i$ with $x_i$ 
being randomly and uniformly distributed in $[0,1]$. We fix the energy (and time) scale by setting $J\!=\!1$. 
%----------------------------------------------------------
\begin{figure}
\includegraphics[width=0.9\columnwidth]{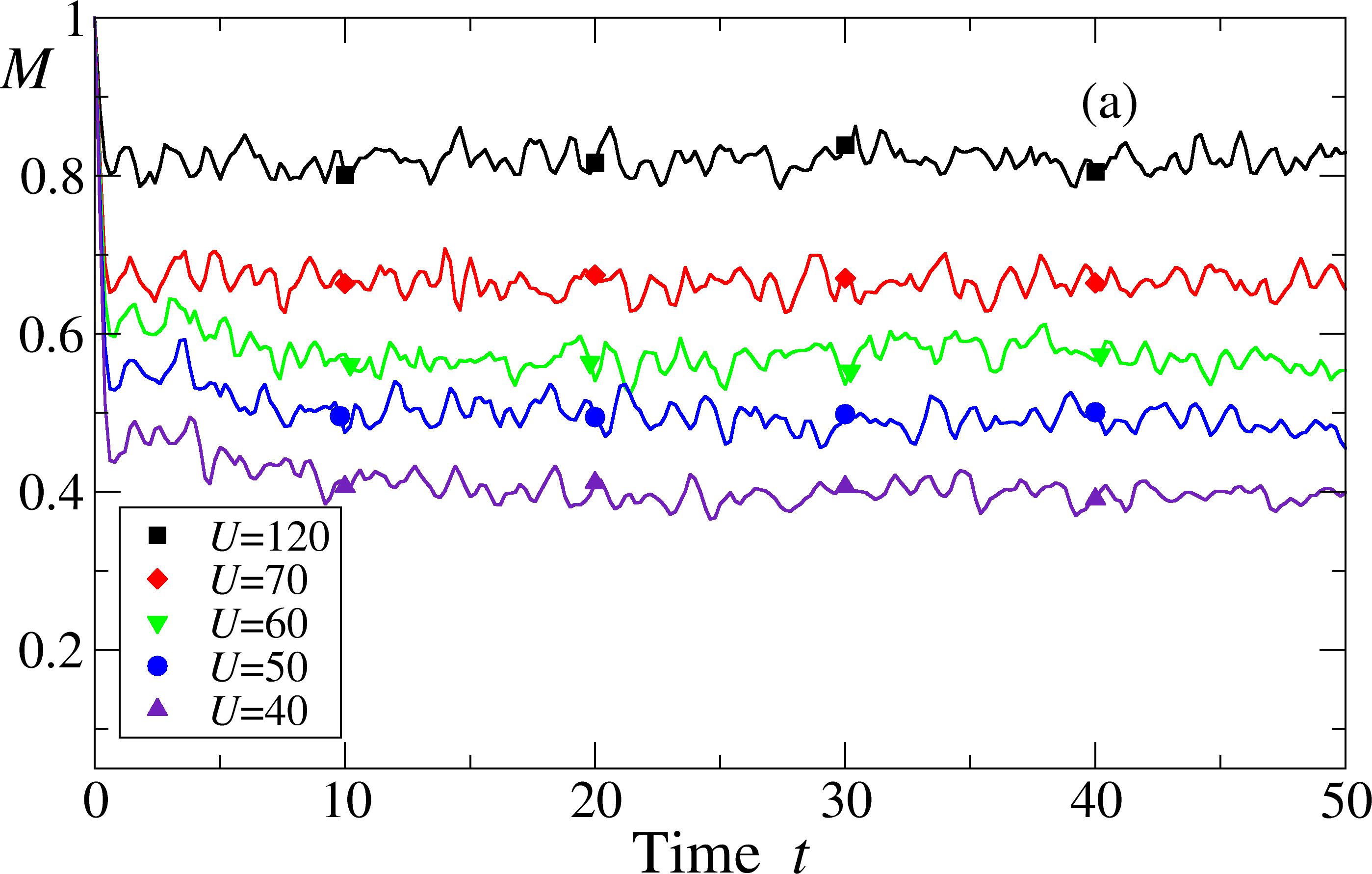}
\includegraphics[width=0.9\columnwidth]{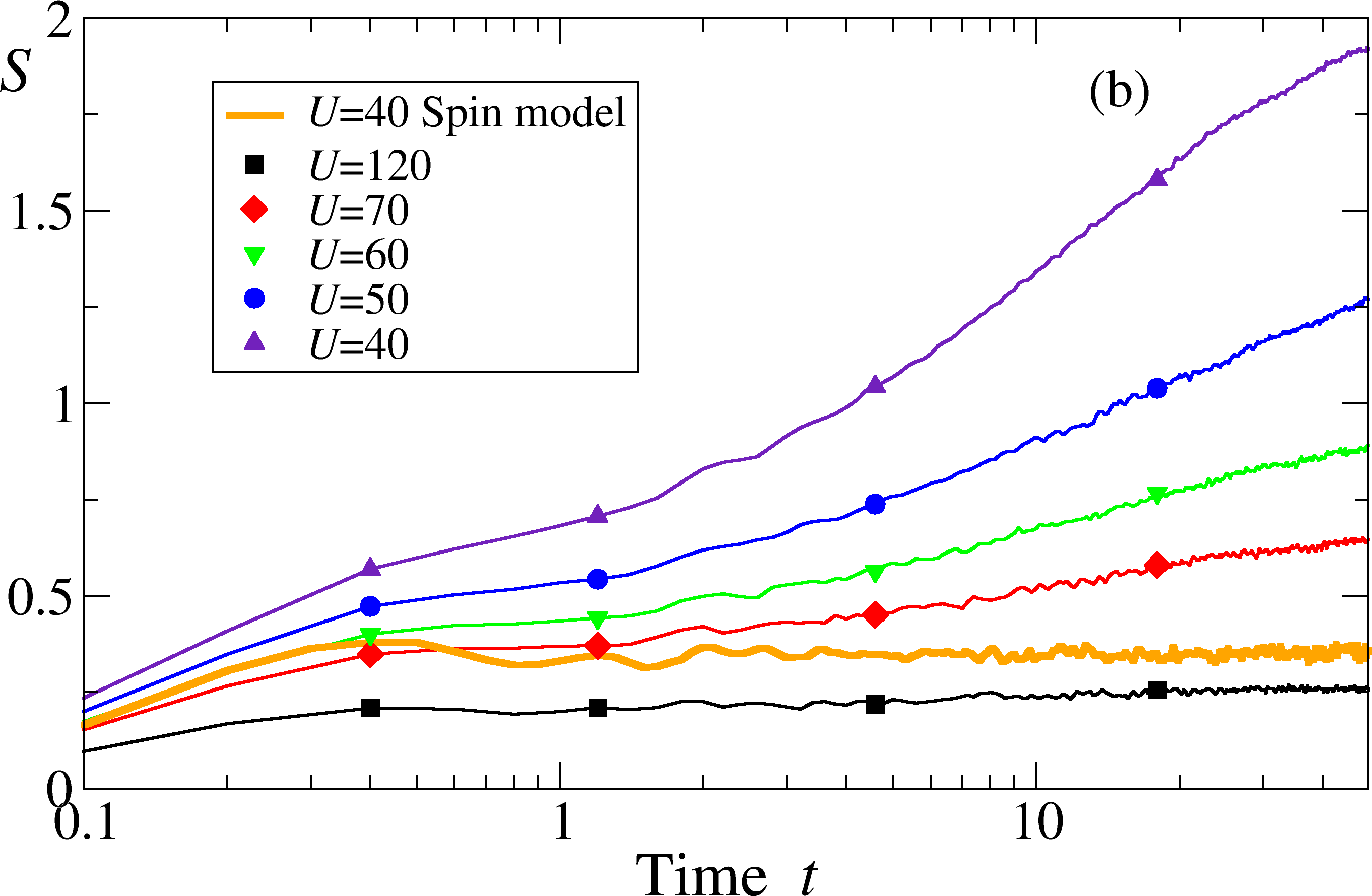}
%\includegraphics[width=0.9\columnwidth]{sdzfig1a.png}
%\includegraphics[width=0.9\columnwidth]{sdzfig1b.png}
%\vspace{-2.0cm}
\caption{ 
\label{bos1}(color online) Many-body localization for a 1D Bose-Hubbard model with random interaction strength. An initial state with a density wave profile 
is temporally propagated using a tDMRG algorithm. 
(a): The magnetization $M$ (initially unity) rapidly decays to a non-zero quasi-stationary value, a clear-cut proof of absence of thermalization, i.e. of MBL. 
(b): The corresponding entropy of entanglement $S$ grows logarithmically with time $t$ after the initial transient. 
The magnetization increases with the disorder strength $U$ correlating with a slower increase
of the entanglement entropy. Results are averaged over 10 disorder realizations, for system size $L=60$ and $N=90$ bosons. 
The additional curve, denoted ``$U=40$ Spin model'', is 
for the spin model discussed in the text. This simplified model exhibits Anderson localization and gives rise to 
a magnetization comparable to the one of the full model at large $U$, see Fig.~\ref{scal1}, but the entanglement entropy
saturates at long time, emphasizing the difference between single particle and many-body localization.
}
\end{figure}
%----------------------------

Could this system reveal MBL? To check this, let us use an experimental approach \cite{Schreiber15} 
and study the temporal dynamics and possible thermalization of some specially prepared state at fixed {particle} density $N/L\!=\!1.5$. 
We prepare the system in a density wave (DW) 
state with odd sites being singly and even sites doubly occupied. We define the magnetization as $M=3(N_e-N_o)/(N_e+N_o)$ where $N_e$ ($N_o$)
corresponds to global populations of even (odd) sites. 
The thermalization hypothesis  suggests that the magnetization, originally equal to unity, 
 would  decay to zero in time (after disorder averaging). 
Yet the numerical results obtained using a home-made implementation of the tDMRG algorithm \cite{Vidal03,Vidal04,Zakrzewski09,Schollwoeck11}) 
suggest otherwise: in Fig.~\ref{bos1}(a), 
at long times $t$, the magnetization $M$ fluctuates about a non-zero mean value that depends on $U$.
Thus the system, despite strong interactions, remembers about the initial state, i.e. does not thermalize. 
This shows that random interactions partially inhibit transport between neighboring sites.

A characteristic feature predicted for the MBL is the logarithmic in time growth of the entanglement entropy. 
To be specific consider $S=-\sum \lambda_i \ln \lambda_i$ where $\lambda_i$ are the Schmidt decomposition coefficients
when tracing out a part of the system (since we can arbitrarily split the 1D chain into two parts such different 
splittings allow for an additional averaging). Indeed, our results display such a logarithmic growth of $S$ as shown in Fig.~\ref{bos1}(b). 

\begin{figure}
\includegraphics[width=0.96\columnwidth]{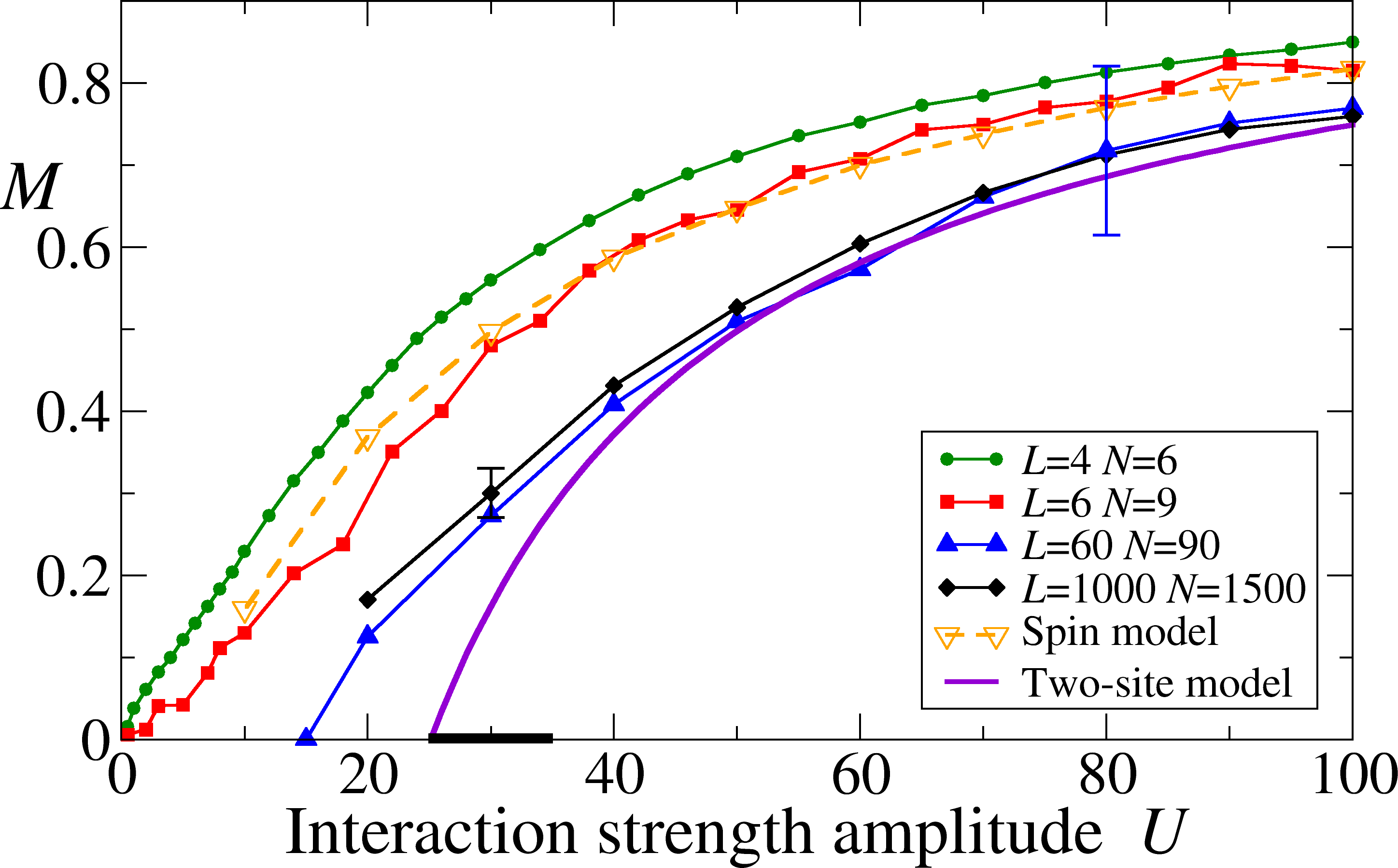}
%\vspace{-2.0cm}
\caption{ \label{scal1}(color online) Quasi-stationary magnetization, $M$, versus $U$ for different system sizes $L$  
indicated in the figure. Data are averaged over time for long times $t\!\in\![10,40]$ to remove the oscillatory behavior visible in Fig.~\ref{bos1}(a). 
Data for small system sizes, obtained from exact diagonalization, are averaged over several hundred realizations of disorder. 
Results for $L\!=\!60$ {(resp. $L\!=\!1000$)} are obtained using the tDMRG algorithm and are averaged over 20  {(resp. 4)} realizations. {For readability, the error bars (one standard deviation) are shown for a single $U$ value, but are very similar for all $U$ values. The quasi-coalescence of the $L=60$ and $L=1000$ results indicate the absence
of finite-size effects.}
The simplified spin model (dashed curve) discussed in the text,
slightly overestimates the magnetization, but catches correctly the asymptotic
behavior at large $U.$ The two-site model, with its analytic prediction $M\approx 1-8\pi/U$ (see text), reproduces quantitatively the results at large $U.$ The black bar in the range $U\!\in\![25,35]$ indicates the region where the transition to MBL occurs - see discussion in the text.
}
\end{figure}

%----------------------------------------------------------
\begin{figure}
\includegraphics[width=0.96\columnwidth]{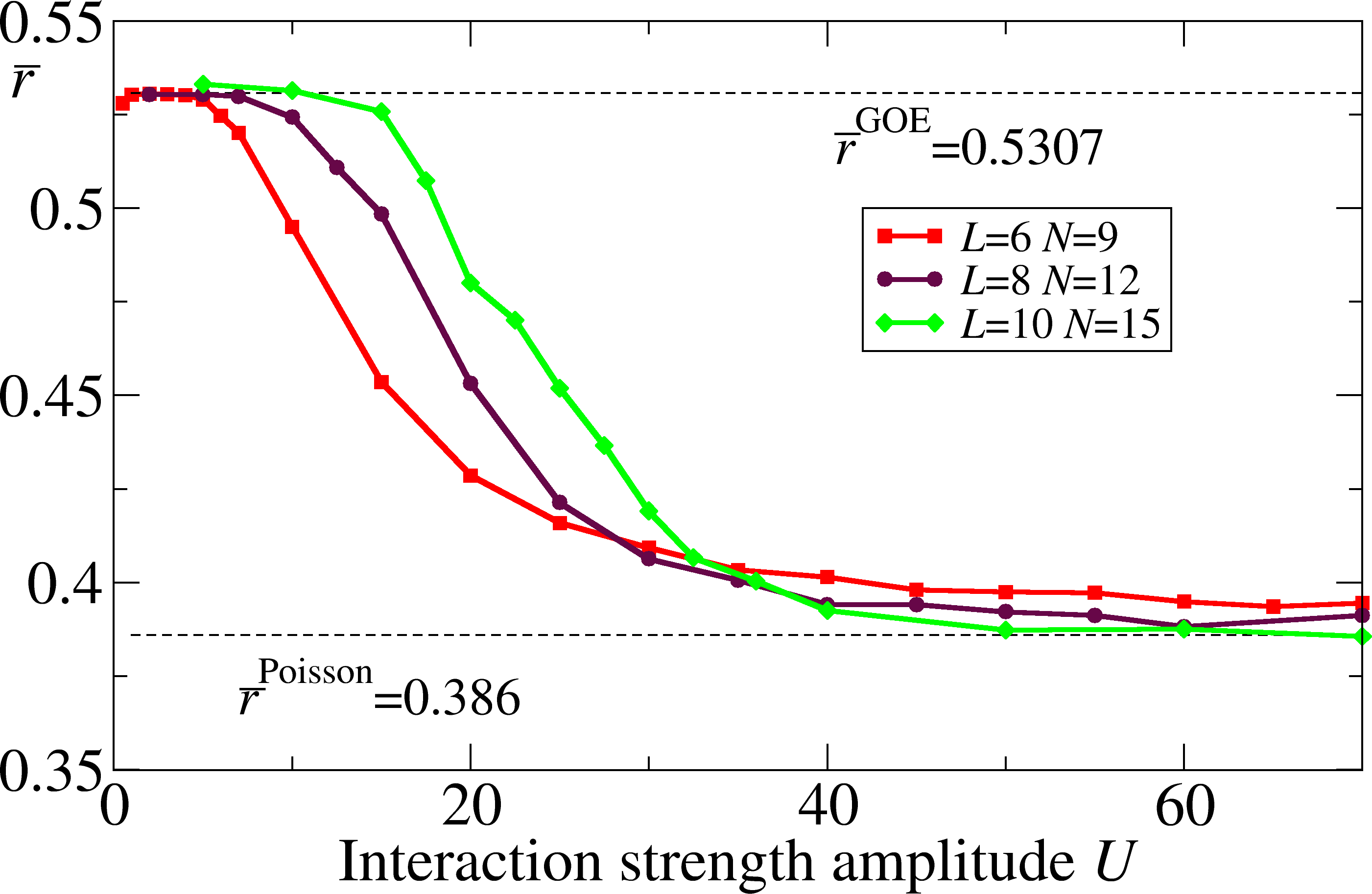}
%\vspace{-2.0cm}
\caption{ \label{rforb}Average ratio of adjacent energy gaps  
$\bar r$ as a function of the interaction strength amplitude $U$ for different system sizes $L$ and number of bosons $N$ 
with a fixed {density} 3/2. 
Data are averaged over many disorder realizations. Only the energy range where eigenstates significantly overlap with 
the initial state is taken into account to facilitate a relevant comparison of time and spectral information. {The dashed horizontal lines  are the GOE and Poisson predictions.}
}
\end{figure}
%----------------------------

Fig.~\ref{scal1} shows the quasi-stationary long time magnetization $M$ as a function of the disorder strength $U$. 
The non-zero magnetization, a characteristic feature of the lack of thermalization,  depends on the system size $L$ 
and can be seen to occur already for small $U$. With increasing $L,$ the magnetization shifts to the right 
converging in the large $L$ limit. Unfortunately, reliable tDMRG calculations for small $U$ cannot be performed for
sufficiently long times due to the growth of the entanglement. That indicates lack of localization for small $U$. 
On the localized side, the logarithmic in time growth of the entropy is observed for $U$ larger than $\sim 30$ only. In the intermediate range $U\!\in \![20,30],$ the numerical results tend to indicate
	a non-zero magnetization at long time, accompanied
	by a rapid, power-like growth of the entropy of entanglement, resembling the observations of \cite{Luitz16,Luitz-rev16} for the XXZ model in the delocalized phase.  
The lack of reliable numerical results for long times and small $U$ 
unfortunately prevented us from determining whether the transition is a smooth cross-over or a phase transition in 
the thermodynamic limit; such an information would be a very important characteristic property of MBL.  The numerical range where the transition occurs is denoted by the black bar in Fig.~\ref{scal1}.

While the observed MBL arises solely due to random interactions and seems to have a non-perturbative character,
its main properties can be understood from a simple microscopic model.
For large $U$, the energy region where the initial state  $|1,2,1,2,1..\rangle$ lives is dominated
by states having the same number of single and double occupations in a random order 
(and additionally preserving the total number of atoms). 
Indeed, moving from a 1,2 configuration to a 2,1 configuration costs no energy on average, while moving to a 0,3 
configuration costs about $U.$ We thus consider a simplified model, where 
the occupation numbers of all sites are either 1 or 2 only. 
This problem maps to a spin model (XX Heisenberg model for a spin-1/2 chain) in a random magnetic
field, see Hamiltonian (1) in \cite{Bardarson12} or Hamiltonian (2) with $\Delta=0$ in \cite{Znidaric08}.
A Wigner-Jordan transformation maps the latter system  onto noninteracting fermions with random diagonal disorder that 
exhibit Anderson localization.  Restricting occupations accordingly, we obtain in our tDMRG
calculations a long time magnetization comparable to the one of the full model, see Fig.~\ref{scal1}, as well
as a rapid saturation of the entropy of entanglement, see Fig.~\ref{bos1}(b).

For very large $U,$ a simpler ''two-site`` model can be built. On-site energies typically differ by much more than
$J$, inhibiting transport. It is only when two neighboring sites have
accidentally on-site energy difference of the order of $J$ that a
significant transfer can take place. This happens with probability
$\propto J/U.$
Consider two states on nearby sites having, respectively, 
occupations $|2,1\rangle$ and  $|1,2\rangle,$ and random interaction energies $U_1$ and $U_2$.  Averaging over
the random disorder as well as over time oscillations, one can compute the magnetization  $M\approx 1-8 \pi J/U$ for
large $U$ which reproduces well the numerical observations. 

In view of these results, it is {also} interesting to  analyze the 
statistical properties of the energy spectrum. Indeed, in the delocalized/ergodic phase, one expects the Gaussian Orthogonal
Ensemble (GOE) of Random Matrix Theory to be relevant, especially
with linear repulsion between neighboring levels, corresponding to complete
delocalization in the Hilbert space. In contrast, the MBL phase is expected to lead to the absence of level repulsion
and Poisson level statistics. 
A simple indicator is
the average ($\bar r$) ratio between the smallest and the largest adjacent energy gaps:
$r_n = \min[\delta^E_n ,\delta^E_{n-1}]/ \max[\delta^E_n,\delta^E_{n-1}],$ with $\delta^E_n=
E_n - E_{n-1},$ and $E_n$ is the ordered list of energy levels \cite{Oganesyan07}. 
In the ergodic (resp. MBL) phase, one expects $\bar r$ to be close to {the GOE} value
{$\bar r^\mathrm{GOE}\approx0.5307$ (resp. $\bar r^\mathrm{Poisson}=2\ln 2 -1\approx0.386$)}
\cite{Atas13}. 

{The localized/ergodic dynamics depends on energy \cite{Luitz15}. For example, we have checked that an initial state $|0,3,0,3...\rangle$ with the same 1.5 particle density leads similarly do a decay of the magnetization with time, but displays MBL for a significantly smaller $U$ value. The statistical properties of the energy levels are also likely to depend on energy}
 so it is important to specify the energy range.
While \cite{Tang15,Mondaini15} used arbitrarily the central part of the spectra in their  
 study of $\bar r$, we choose the vicinity of the energy of our DW initial state.
This is also the region of significant overlaps between eigenstates and the initial state. 
The results for different system sizes  are presented in Fig.~\ref{rforb} providing {another} evidence for the transition to MBL for sufficiently large $U$. 
Crossings of data for numerically accessible system sizes do not allow 
to precisely pin down the transition/crossover point, which lies probably around $U\approx 30,$ slightly larger than { -- but compatible with -- } {the} magnetization data.

An intriguing possibility \cite{Luitz16,Luitz-rev16,Torres-Herrera16} is the possible existence of a delocalized but non ergodic phase below the critical point. Although the numerical simulations are very difficult in this region because of the rapid increase of the entanglement entropy,
the data shown in Fig.~\ref{scal1} for large system size tend to show that there is a non-zero magnetization at long time in the $U\!\in\![20,30]$ range, while the statistical properties in Fig.~\ref{rforb} tend to show that this takes place in the delocalized regime. We thus conclude that our results are in favor of the existence of a non-ergodic delocalized phase, although we admit that further work is required to confirm this observation.

%----------------------------------------------------------
\begin{figure}
\includegraphics[width=0.99\columnwidth]{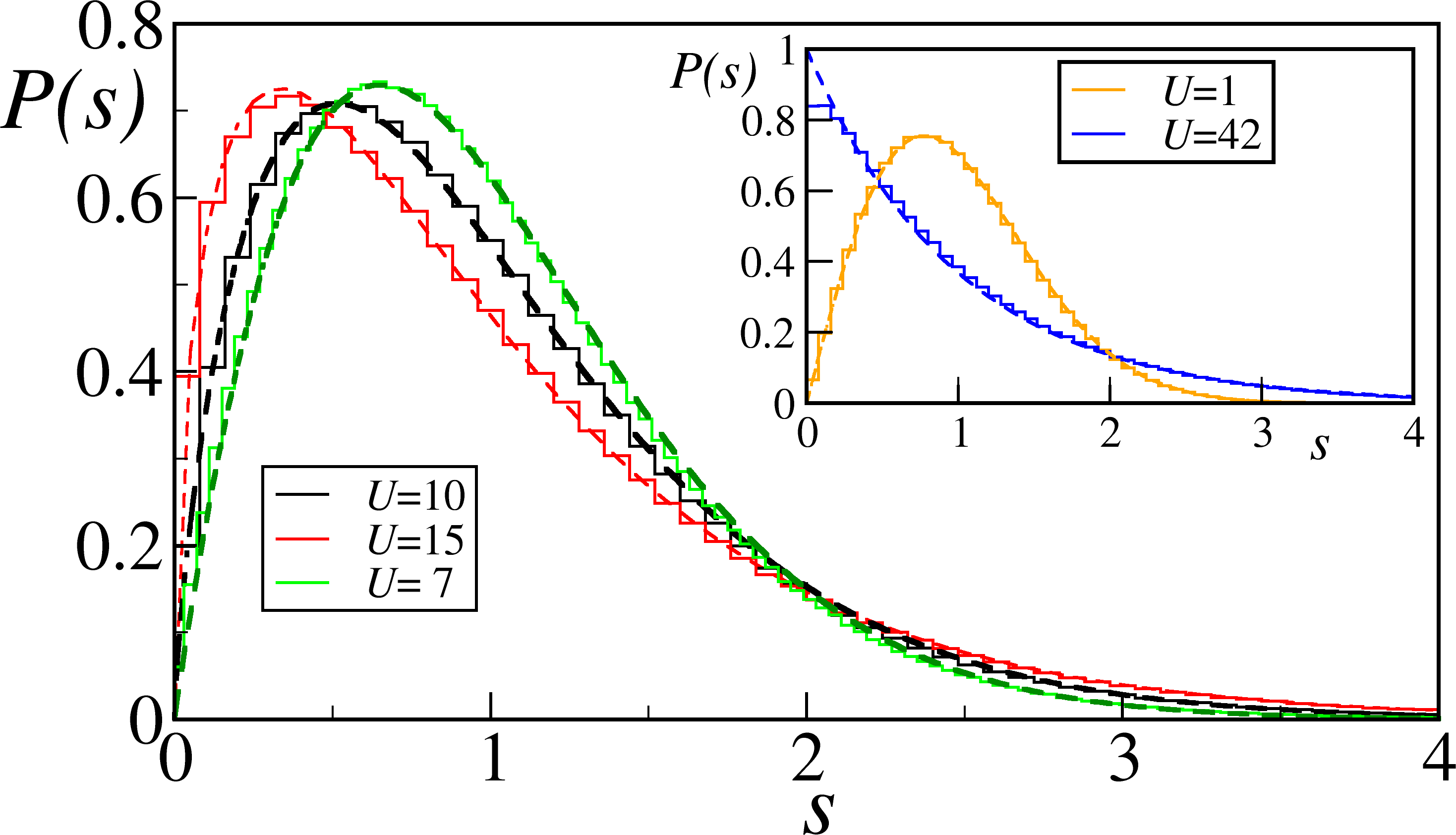}
%\vspace{-2.0cm}
\caption{ \label{spacing} (color online) Level spacing distributions for $N=9$ particles on $L=6$ sites with open boundary conditions after averaging over several realizations of  disorder {with strength $U$}. 
{The $U=15$ data is accurately described by the semi-Poisson distribution $P(s)\propto s^\beta\exp(-(1+\beta)s)$ with $\beta\approx 0.508$, as expected in the critical region. At lower $U$ (7 and 10) there is a transition towards the GOE distribution, where the data
are well reproduced by a $P(s) \propto s^{\beta} \exp(-C_2 s^{2-\gamma})$ distribution proposed in~\cite{Serbyn16}.} 
The inset shows two limiting cases: the small (resp. large) $U$ distribution is well reproduced by the  GOE (resp. Poisson) prediction.
}
\end{figure}
%---------------------------

The analysis of {level statistics} can be carried further. This requires to perform the standard
``unfolding'' (using a polynomial fit) of the
energy spectrum, obtaining level sequences of unit mean spacing [In contrast, the $\bar r$ statistics does not require such an unfolding \cite{Atas13}].
This allows us to study spacing distributions for different $U$ values and probe the crossover region more carefully \cite{Serbyn16,Garcia-Garcia}.  
 For small $U,$ a  
very good agreement with the GOE prediction is observed in accordance with the $\bar r $ value, compare with Fig.~\ref{spacing}.  
With increasing $U$ we observe a transition of spacing distributions towards a typical shape expected for localized distributions.
The study of the so-called intermediate statistics has proved useful in the context of single particle localization~\cite{Montambaux98}.
{Around the critical point, the data are well described by the semi-Poisson family describing the spacing distribution:  $P(s) \propto s^\beta\exp(-(\beta+1)s)$ (with $\beta$ a real parameter), as shown in Fig.~\ref{spacing} for $U=15$ and tend to the Poisson distribution (with
	$\beta=0$) for very large $U.$
	On the delocalized side, for $U=7$ and 10 in Fig.~\ref{spacing}, the situation is a bit more complicated,
	with an intermediate regime well described by the distribution  $P(s) \propto s^{\beta} \exp(-C s^{2-\gamma})$ proposed in \cite{Serbyn16} 
	in the context of the XXZ spin chain {(a similar behavior is observed in~\cite{Garcia-Garcia})}. We speculate that it could be related to the transition from a non-ergodic delocalized phase to an ergodic one.}

In conclusion we have shown that many body localization may be observed in one-dimensional systems in optical lattices,
under realistic experimental conditions when the disorder is due to random interactions (with no disorder in the chemical potential). 
The virtue of our model is that it does not show localization without interactions:
 the MBL effect observed is inherently and solely due to interactions and is not a small perturbation {of} single particle physics.
 We believe that this supports the 
 idea that MBL is robust (as suggested by many-body strongly coupled systems without 
 disorder \cite{Yao2014,Roeck14,Grover14,Schiulaz15,Goold2015}).

This research was performed within 
project   No.2015/19/B/ST2/01028  financed by  National Science Centre (Poland). Support by PL-Grid Infrastructure, EU via project
QUIC (H2020-FETPROACT-2014 No. 641122)  and the Polish-French bilateral program POLONIUM (Grant 33162XA) is  also acknowledged.

\bibliography{references_v04}

\end{document}